\documentclass[conference]{IEEEtran}
\IEEEoverridecommandlockouts
\usepackage{cite}
\usepackage{amsmath,amssymb,amsfonts}
\usepackage{algorithmic}
\usepackage{algorithm}
\usepackage{graphicx}
\usepackage{textcomp}
\usepackage{xcolor}
\usepackage{mathrsfs}
\usepackage{multirow}
\def\BibTeX{{\rm B\kern-.05em{\sc i\kern-.025em b}\kern-.08em
    T\kern-.1667em\lower.7ex\hbox{E}\kern-.125emX}}
\begin{document}

\newcommand{\CAREPL}[2]{\textcolor{red}{#1}\textcolor{green}{#2}}
\newcommand{\CANOTE}[2]{\textcolor{orange}{#1}{\footnote{\textcolor{orange}{#2}}}}

\title{Reliable Long Timescale Decision-Directed Channel Estimation for OFDM System}

\author{\IEEEauthorblockN{Xun Wang\IEEEauthorrefmark{1},
		Xin Xie\IEEEauthorrefmark{2},
		Cunqing Hua\IEEEauthorrefmark{2}, 
		Jianan Hong\IEEEauthorrefmark{2},
		Pengwenlong Gu\IEEEauthorrefmark{3}
		}
		\IEEEauthorblockA{wangx6258@sjtu.edu.cn,
		xiexin\_312@sjtu.edu.cn,
		cqhua@sjtu.edu.cn,
		hongjn@sjtu.edu.cn,
		pengwenlong.gu@inria.fr,
		}
		\IEEEauthorblockA{
		\IEEEauthorrefmark{1}School of Engineers Paris, Shanghai Jiao Tong University, Shanghai, China,\\
		\IEEEauthorrefmark{2}School of Cyber Science and Engineering, Shanghai Jiao Tong University, Shanghai, China,\\
		\IEEEauthorrefmark{3}Inria, France
	}
}

\maketitle

\begin{abstract}
Decision-directed channel estimation (DDCE) is one kind of blind channel estimation method that tracks the channel blindly by an iterative algorithm without relying on the pilots, which can increase the utilization of wireless resources. However, one major problem of DDCE is the performance degradation caused by error accumulation during the tracking process. In this paper, we propose a reliable DDCE (RDDCE) scheme for an OFDM-based communication system in the time-varying deep fading environment. By combining the conventional DDCE and discrete Fourier transform (DFT) channel estimation method, the proposed RDDCE scheme selects the reliable estimated channels on the channels that are less affected by deep fading and then estimates the channel based on the selected channels by an extended DFT channel estimation where the indices of selected channels are not distributed evenly. Simulation results show that RDDCE can alleviate the performance degradation effectively, track the channel with high accuracy on a long time scale, and has good performance under time-varying and noisy channel conditions.
\end{abstract}


\section{Introduction}

Orthogonal frequency-division multiplexing (OFDM) \cite{OFDM} has been widely adopted in many universal digital communication systems (e.g., LTE\cite{3gpp_36_211}, NR\cite{3gpp_38_211}, WiFi\cite{IEEE_802_11}) to address the inter-symbol inference (ISI) problem caused by multi-path propagation in the wireless channel. By using the excellent feature of DFT (Discrete Fourier Transform), the complicated frequency-selective fading can be transformed to parallel flat-fading in each sub-carrier in an OFDM system\cite{tse_viswanath_2005} whereby a relative simple equalization scheme in the frequency domain is sufficient to extract the information from the received symbols. 

On top of that, a channel estimation process is necessary to obtain knowledge of channel fading before carrying out the equalization process. Pilot-based channel estimation is one kind of channel estimation method that is widely used, whereby the transmitter inserts a special kind of symbol named reference symbols or pilots between the data symbols, which are defined in the protocol and known by both the transmitter and receiver. After the receiver receives the signal, it will first estimate the channel based on pilots using some channel estimation algorithms, such as Least Square (LS) or Minimum Mean Square Error (MMSE)\cite{MMSE}, and then equalize the channel to demodulate the data symbols. 
Pilot-based channel estimation performs well in many cases. However, one drawback is that the pilots occupy the communication bandwidth and reduce the utilization of wireless resources, especially under bad communication conditions, more pilots are needed to recover the data symbols correctly.

The blind channel estimation schemes have been proposed to address this problem, which do not insert pilots between data symbols and estimate the channel blindly. Decision-directed channel estimation (DDCE) is one kind of blind channel estimation method that utilizes the preamble signal to obtain the initialized channel estimation and then performs an iterative algorithm to track the time-varying channel. 
DDCE does not require extra bandwidth except for the preamble signal, which is negligible compared with the total amount of transmitted information, so it has higher bandwidth utilization than the pilot-based channel estimation method. However, since DDCE is an iterative algorithm, any estimation error during the iteration will accumulate over time and cause degradation of estimation performance, which is one of the main challenges of DDCE.

In order to improve the performance of DDCE methods, many research efforts have been made in the past few years.
In \cite{fadingenvelope}, DDCE is combined with the channel prediction method to address the problem of DDCE under the fading condition. In ordinary situations, the estimator tracks the channel by DDCE. But in a deep fading, the estimator will rely upon a predicted value rather than the result returned by DDCE.
In \cite{cooperativesubcarrier}, considering that less error will occur if the modulation method of lower code rate is applied, DDCE is performed on the cooperative channels where data symbols are modulated by M-ary phase shift keying (MPSK) and M-ary amplitude phase shift keying (MASK).
In \cite{gamma}, a recursive filter is used to provide reliable feedback information in DDCE for OFDM systems with high velocities. The application of Deep Learning in DDCE has been studied by \cite{dlddce}, whereby deep neural network $k$-step predictor-based DDCE is applied for a multiple-input multiple-output (MIMO) communication system in highly dynamic vehicular environments. Although these methods can alleviate the error accumulation of DDCE, the performance degradation of DDCE is still severe under time-varying deep fading channels on a long time scale. Therefore, DDCE is usually applied in the scenario where the channel is nearly static or in a very short period and will be reinitialized by inserting pilots again for a few frames.

In this paper, we propose a reliable DDCE (RDDCE) method that combines the conventional DDCE method with DFT channel estimation to address the performance degradation problem of DDCE. Specifically, the set of reliable channels that are less affected by deep fading are selected based on the preliminary estimation results of DDCE, and then the complete channel on all channels are estimated by the extended DFT channel estimation method.
Simulation results show that our method can filter out the misjudged channel effectively and avoid error accumulation during the iteration. On a long timescale, our method can track the channel with high accuracy.

The remainder of the paper is organized as follows. In Section \ref{systemmodel}, we introduce the OFDM wireless communication system model under a time-variant frequency selective fading channel, give a brief description of related methods, and form the problem. In Section \ref{method}, we present the RDDCE channel estimation method. In Section \ref{results}, we provide simulation results to evaluate the performance of RDDC in different scenarios. Section \ref{conclusion} concludes this paper.

\section{System Model}\label{systemmodel}
In this paper, we consider an OFDM communication link between a transmitter and a receiver. For this link, the channel is a quasi-static multipath channel that follows a frequency-selective Rayleigh fading and is subject to a circularly symmetric complex Gaussian (CSCG) distribution $\boldsymbol{W}_t\sim\mathcal{C}\mathcal{N}(0,\sigma^2)$. 
"Quasi-static" implies that the channel taps are constant over one OFDM symbol and change slowly between consecutive OFDM symbols with respect to the OFDM symbol duration.

The channel at a time instance $t$ could be represented by $\boldsymbol{H}_t\in\mathbb{C}^{N_c \times 1}$, where $N_{c}$ is the number of channels in one OFDM symbol. For a transmitted signal $\boldsymbol{X}_{t}[k]$ on the $k$th channel, the corresponding signal $\boldsymbol{Y}_{t}[k]$ received at the receiver can be expressed as 
\begin{equation}\label{eq:signal}
\boldsymbol{Y}_{t}[k]=\boldsymbol{H}_{t}[k]\boldsymbol{X}_{t}[k]+\boldsymbol{W}_{t}[k]
\end{equation}
where $k\in[\![0,N_{c}-1]\!]$.

We assume that all path delays are integer multiplies of sample period $T_{s}$, and 
there exists a maximum channel delay $\tau_{max}$ that can be known a priori.


In the following, we briefly introduce two classic channel estimation methods, which will be referred to in our method.

\subsection{Decision-Directed Channel Estimation}
\label{decision-directed}

DDCE is one kind of iterative channel estimation algorithm, which utilizes the previously estimated channels to demodulate data symbols of the current time instance and then uses the data symbols as pilots to estimate the current channel. Considering the simplest situation, only the estimated channel of last time instance $\hat{H}_{t-1}$ is applied, then the DDCE scheme consists of the following three steps:
\begin{enumerate}
	\item  Equalization: $\hat{\boldsymbol{X}}_t = f_{eq}(\boldsymbol{Y}_t, \hat{\boldsymbol{H}}_{t-1})$;
	\item Demodulation: $\bar{\boldsymbol{X}}_t = f_{dec}(\hat{\boldsymbol{X}}_t)$;
	\item Estimation: $\hat{\boldsymbol{H}}_t = f_{est}(Y_t, \bar{\boldsymbol{X}}_{t})$.
\end{enumerate}

Specifically, in step one, the channel $\hat{\boldsymbol{H}}_{t-1}$ is equalized to get estimated symbols $\hat{\boldsymbol{X}}_t$, then the symbols are demodulated by the hard decision function $f_{dec}$ to get $\bar{\boldsymbol{X}}_t$ in the second step. Finally,  the channel is updated based on $\boldsymbol{Y}_t$ and $\bar{\boldsymbol{X}}_t$ by estimation function $f_{est}$.

One major problem of DDCE is the demodulation error in the second step. Under time-varying channel conditions,  this error will occur as we equalize the channel of the last time instance rather than that of the current one in the first step. Since DDCE is a recursive algorithm, the estimation error will be accumulated and lead to performance degradation.
Although some solutions have been proposed to alleviate the performance degradation, which is still severe for a long time duration. 

\subsection{DFT Channel Estimation}

DFT channel estimation is a very common pilot-based channel estimation method \cite{cesurvey}. Suppose that at time $t$, we have the estimated channel $H_{t,p}$ in the position of pilots, then the completed channel $H_t$ can be estimated by the following steps:

\begin{enumerate}
	\item Perform IDFT on $\boldsymbol{H}_{t,p}$ to get channel impulse response $\boldsymbol{h}_{t}$:
	\begin{align}
		\boldsymbol{h}_t = \boldsymbol{F}^{-1}\boldsymbol{H}_{t,p}
	\end{align}
	where
	\begin{align}
		\boldsymbol{F}[i,j] = exp(-\frac{2\pi ij}{N_c}), \quad i,j\in[\![0, N_c - 1]\!]
	\end{align}
	\item Eliminate the values of $h_{t}$ which is out of the maximum path delay $\tau_{max}$;
	\item Pad zeros after $\boldsymbol{h}_{t}$ until its length equals to $N_c$;
	\item Perform DFT to the padded $\boldsymbol{h}_{t}$ to get the estimated channel $\hat{\boldsymbol{H}}_{t}$:
	\begin{align}
		\hat{\boldsymbol{H}}_t = \boldsymbol{F}\boldsymbol{h}_{t}
	\end{align}
\end{enumerate}


\section{Reliable Decision-Directed Channel Estimation Scheme}\label{method}

In this paper, we propose RDDCE, a reliable DDCE method that combines the conventional DDCE and DFT channel estimation. It attempts to filter out channels with estimation error dynamically, and thus avoid the accumulation of estimation errors on a long timescale.

\begin{figure}
	\centerline{\includegraphics[scale=0.4]{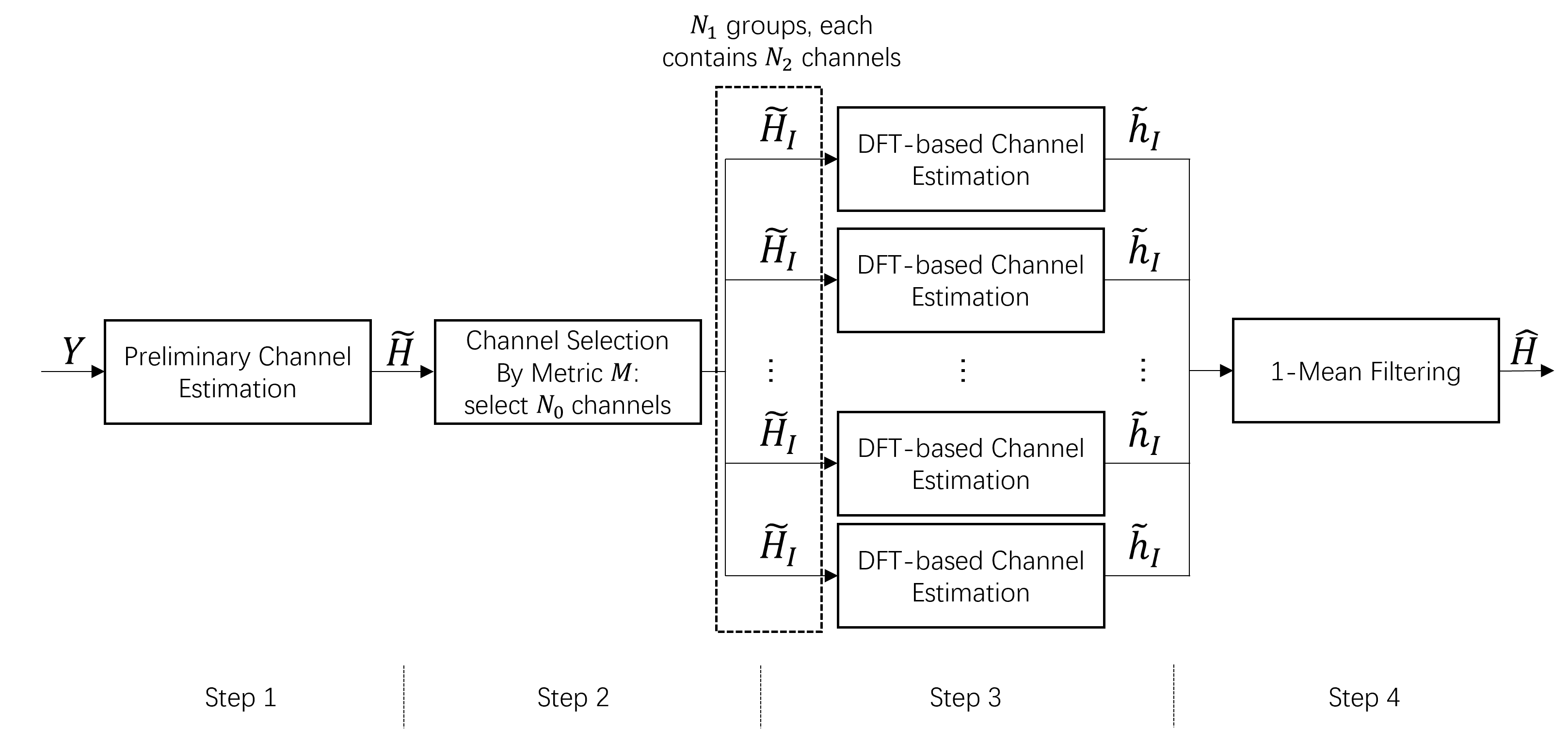}}
	\caption{Channel estimation pipeline of RDDCE constituted by four steps: preliminary estimation, channel selection, DFT-based channel estimation by group and 1-mean filtering.}
	\label{overview}
\end{figure}

The pipeline of our RDDCE scheme is shown in Fig.\ref{overview}, which consists of four steps: preliminary channel estimation, channel selection, DFT channel estimation by group, and 1-mean filtering. In the preliminary estimation step, the channel is estimated using the DDCE scheme that has been explained in Section~\ref{decision-directed}. In the channel selection step, we select $N_0$ channels with a lower estimation error probability from the total preliminary estimated channel. 
In the DFT channel estimation step, we divide the selected channel into $N_1$ groups and perform DFT channel estimation for each group to get their channel impulse responses (CIR), which can be considered as observations of the actual CIR. 
Finally, in the 1-mean filtering step, we estimate the actual CIR based on the observations and perform DFT to get the final estimated channel.



\subsection{Channel selection} \label{picking}
In this step, we select the channels with low estimation error probability.
According to \cite{fadingenvelope}, the channel estimation error may occur in different channels and will happen with higher probability at the locations with
weak $\tilde{\boldsymbol{H}}_t$.
Therefore, a threshold is proposed to filter out the channels with weak $\|\tilde{\boldsymbol{H}}_t\|_2$. But $\|\tilde{\boldsymbol{H}}_t\|_2$ is not highly related to the estimation error.
Specifically, from \eqref{eq:signal}, if the noise is ignored, we can recover $\boldsymbol{X}_{t}$ based on $\hat{\boldsymbol{H}}_{t-1}$ and $\boldsymbol{Y}_{t}$. The $L_2$ distance between the estimated symbol $\hat{\boldsymbol{X}}_{t}[k]$ and the transmitted symbol $\boldsymbol{X}_{t}[k]$ is given by: 
\begin{align} 
	\| \hat{\boldsymbol{X}}_{t}[k] - \boldsymbol{X}_{t}[k]\| &= \| \frac{\boldsymbol{Y}_{t}[k]}{\hat{\boldsymbol{H}}_{t-1}[k]} - \boldsymbol{X}_{t}[k]\| \\
	&= \| \boldsymbol{X}_{t}[k] \|\frac{\| \boldsymbol{H}_{t}[k] - \hat{\boldsymbol{H}}_{t-1}[k] \|}{ \| \hat{\boldsymbol{H}}_{t-1}[k] \|} \label{eq:estimationerror}
\end{align}
It can be seen that the bigger the distance, the higher the estimation error probability. $\| \boldsymbol{X}_{t}[k] \|$ is constant, so the distance is determined mainly by the second term.

Based on \eqref{eq:estimationerror}, we propose a more effective metric $M$ to filter out unreliable channels, which is defined in \eqref{def_M}. We choose $N_{0}$ channels with the smallest $M$.

\begin{align} \label{def_M}
	M(t,k) = \frac{\| \tilde{\boldsymbol{H}}_{t}[k] - \hat{\boldsymbol{H}}_{t-1}[k] \|}{ \| \hat{\boldsymbol{H}}_{t-1}[k] \|}
\end{align}
where $\tilde{\boldsymbol{H}}_{t}$ is the preliminary estimated channel of the current instant, $\hat{\boldsymbol{H}}_{t-1}$ is the final estimated channel of the previous instant. $M$ implies the distance between the estimated symbol $\hat{\boldsymbol{X}}_{t}[k]$ and the transmitted symbol $\boldsymbol{X}_{t}[k]$, which is related directly to the estimation error. The simulation in Section~\ref{sectionperformance} will also prove the effectiveness of $M$.




\subsection{DFT-based channel estimation}
In this step, to estimate 
CIR based on the selected channels, 
we divide the selected channels into $N_1$ groups and each group contains $N_2$ channels. Then we perform DFT channel estimation for each group to get CIR, which can be considered as one of the observations of the actual CIR.

Let $I = \{i_1, i_2, ..., i_{N_{2}}\} \subset [\![1, N_{c}]\!]$ denote the indices of chosen values in one group,  
and 
$\tilde{\boldsymbol{H}}_{I} \in \mathbb{C}^{N_{2} \times 1}$ denotes the channels in this group.
Since $I$ is chosen randomly, we cannot calculate the channel impulse response $\tilde{\boldsymbol{h}_{I}}$ of the group $I$ using the inverse of Fourier transform matrix $F$ as DFT channel estimation usually does. Instead, we calculate $\tilde{\boldsymbol{h}_{I}}$ as follows: 
\begin{align} \label{eq:Htoh}
 	\tilde{\boldsymbol{h}_{I}} = \boldsymbol{F}^{-1}_{I}\tilde{\boldsymbol{H}}_{I}
\end{align}
where $\boldsymbol{F}_{I} \in \mathbb{C}^{N_{2} \times N_2}$ is formed by the lines and columns indexed by $I$ in the Fourier transform matrix $\boldsymbol{F} \in \mathbb{C}^{N_{c} \times N_c}$,
and $\boldsymbol{F}_{I}$ is a Vandermonde matrix, its inversion can be computed quickly by Traub\cite{Traub} or Parker \cite{Parker} inversion algorithm.

\begin{figure}
	\centerline{\includegraphics[scale=0.6]{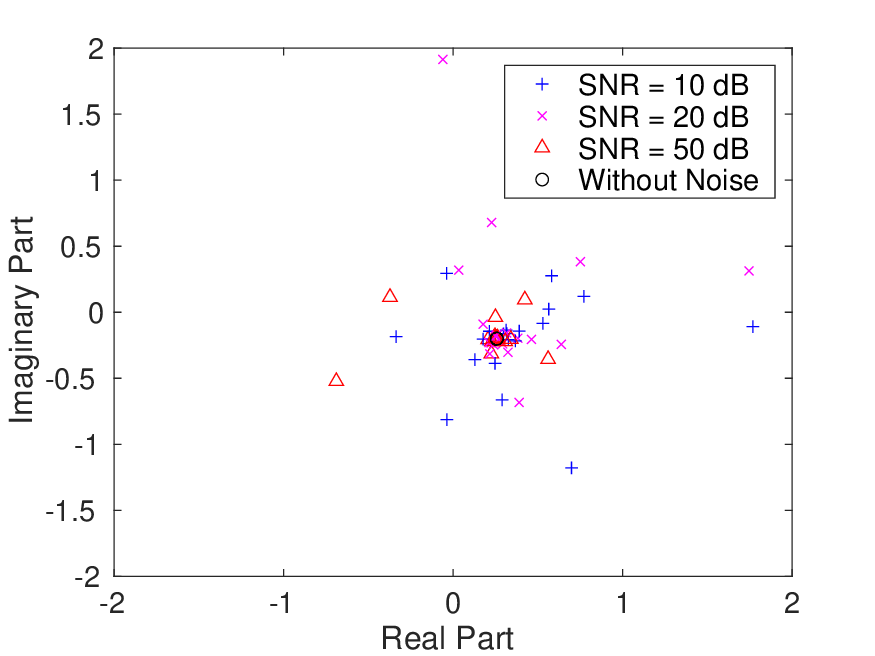}}
	\caption{Scattering of $\tilde{h}_{I}$ under different SNR.}
	\label{diffusion1}
\end{figure}

Based on the simulation, we find that the noise and channel estimation error caused by hard demodulation decision will overwhelm the actual channel impulse responses $\boldsymbol{h}_{actual}$ in $\tilde{\boldsymbol{h}_{I}}$, so which should be eliminated.
Assuming the estimation error can be modelled as an additive Gaussian noise $\boldsymbol{W}\sim\mathcal{C}\mathcal{N}(0,\sigma_2^2)$, then we have the relation between the selected estimated channel $\tilde{\boldsymbol{H}}_{I}$ and actual channel $\boldsymbol{H}_I$:
\begin{align} \label{eq:divideH}
	\tilde{\boldsymbol{H}}_{I} = \boldsymbol{H}_I + \boldsymbol{W}.
\end{align}
Replace $\tilde{\boldsymbol{H}}_{I}$ in \eqref{eq:Htoh} by \eqref{eq:divideH}, then we have:
\begin{align}
	\tilde{\boldsymbol{h}}_{I} &= \boldsymbol{F}^{-1}_{I}\tilde{\boldsymbol{H}}_{I} \\
	&= \boldsymbol{F}^{-1}_{I}(\boldsymbol{H}_I + \boldsymbol{W}) \\
	&= \boldsymbol{h}_{actual} + \boldsymbol{F}^{-1}_{I}\boldsymbol{W} \\
	&\triangleq \boldsymbol{h}_{actual} + \boldsymbol{w},
\end{align}
which implies that $\tilde{\boldsymbol{h}}_{I}$ will scatter from $\boldsymbol{h}_{actual}$ because of 
\begin{align} \label{eq:def_w}
    \boldsymbol{w} = \boldsymbol{F}^{-1}_{I}\boldsymbol{W}.
\end{align}
Fig.~\ref{diffusion1} shows the scattering of $\tilde{\boldsymbol{h}}_{I}$ under different signal-to-noise ratio (SNR) conditions. Each marker in Fig.~\ref{diffusion1} represents the mean value of all taps in one $\tilde{\boldsymbol{h}}_{I}$ corresponding to one $I$. It can be seen that without noise, all black markers assemble in one place $\boldsymbol{h}_{actual}$, and as the noise increases, $\tilde{\boldsymbol{h}}_{I}$ scatters from $\boldsymbol{h}_{actual}$.

\begin{figure}
	\centerline{\includegraphics[scale=0.6]{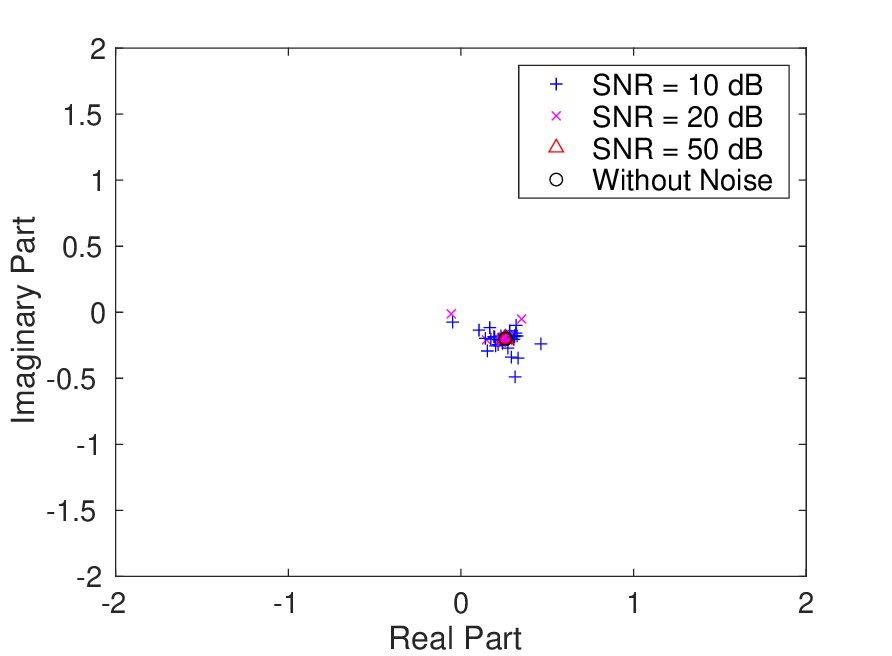}}
	\caption{Scattering of $\tilde{h}_{I}^{'}$ under different SNR.}
	\label{diffusion2}
\end{figure}

To get $\boldsymbol{h}_{actual}$, we need to estimate $\boldsymbol{w}$ and alleviate it. Note that in the situation without noise, $\boldsymbol{h}_{actual}$ only has values in the first $N_{taps}$ taps in $\tilde{\boldsymbol{h}}_{I}$ because the path delay of signal is limited,
so we can estimate the whole $\boldsymbol{w}$ based on the part out of the first $N_{taps}$ taps of $\tilde{\boldsymbol{h}}_{I}$ which does not contain $\boldsymbol{h}_{actual}$.
We set the number of channels included in one group $N_{1} = N_{taps} + N_{w}$, where $N_{w}$ is a number close to $N_{taps}$. Then the last $N_{w}$ values of $\tilde{\boldsymbol{h}}_{I}$ are exactly the last $N_{w}$ values of $\boldsymbol{w}$, which means that $\tilde{h}_{I}$ and $\boldsymbol{w}$ can be divided into two parts:
\begin{align}
	\tilde{\boldsymbol{h}}_{I} = [\boldsymbol{h}_{actual} + \boldsymbol{w}_{1}, \boldsymbol{w}_{2}]
\end{align}
where
\begin{align}
	\boldsymbol{w} = [\boldsymbol{w}_{1}, \boldsymbol{w}_{2}];
    \boldsymbol{w}_{1} \in \mathbb{C}^{N_{taps} \times 1}; \boldsymbol{w}_{2} \in \mathbb{C}^{N_{w} \times 1}
\end{align}
Let us denote the last $N_w$ lines of the inverse of Fourier transform matrix $\boldsymbol{F}_{I}$ as $\boldsymbol{F}_{w}$, based on the function given in \eqref{eq:def_w}, we have:
\begin{align}
	\boldsymbol{w}_{2} = \boldsymbol{F}_{w}\boldsymbol{W}
\end{align}
Therefore, given $\boldsymbol{w}_{2}$, we can estimate $\boldsymbol{w}$ and compensate it by following three steps:
\begin{enumerate}
	\item \label{step1} Estimate noise $\hat{\boldsymbol{W}}$ by the maximum likehood estimation method:
	\begin{align}
		\hat{\boldsymbol{W}}^T = \boldsymbol{F}_{w}^{\dagger}\boldsymbol{w}_{2}^T
	\end{align}
	because $\boldsymbol{W}\sim\mathcal{C}\mathcal{N}(0,\sigma_2^2)$, where $\{\cdot\}^\dagger$ denotes the pseudo inverse matrix,
	\item Estimate the noise $\hat{\boldsymbol{w}}$:
	\begin{align}
		\hat{\boldsymbol{w}} = \boldsymbol{F}^{-1}_{I}\hat{\boldsymbol{W}};
	\end{align}
	\item Denoise:
	\begin{align}
		\tilde{\boldsymbol{h}}_{I}^{'} = \tilde{\boldsymbol{h}}_{I} - \hat{\boldsymbol{w}}
	\end{align}
\end{enumerate}
It should be noted that when $I$ is chosen evenly, $\tilde{\boldsymbol{h}}_{I}^{'}$ equals to $\tilde{\boldsymbol{h}}$ eliminating all taps out of $N_{taps}$ in DFT channel estimation.
The above steps can be considered as an extension of DFT channel estimation in the situation where $I$ is chosen unevenly.
Fig.~\ref{diffusion2} shows the scattering of $\tilde{\boldsymbol{h}}_{I}^{'}$ under different SNR conditions after denoising. Compared with Fig. \ref{diffusion1}, all markers get closer to $\boldsymbol{h}_{actual}$, the scattering has been alleviated effectively.
Repeating the aforementioned steps for all groups, we can get $N_2$ different $\tilde{\boldsymbol{h}}_{I}^{'}$, which can be considered as $N_2$ observations of $\boldsymbol{h}_{actual}$. The next step is to estimate $\boldsymbol{h}_{actual}$ based on these observations.

\subsection{1-Mean Filtering}
In this step, we will get the final estimated channel $\hat{\boldsymbol{H}}$ by finding the position where the majority of estimated CIR $\tilde{\boldsymbol{h}}_{I}^{'}$ are close to.
To filter out the misestimated $\tilde{\boldsymbol{h}}_{I}^{'}$ which are too far from $\boldsymbol{h}_{actual}$, we propose a simple 1-mean algorithm as shown in Algorithm~\ref{alg1}, which is inspired by the $K$-mean algorithm \cite{kmean}.

The 1-mean algorithm starts from the mean value $\tilde{\boldsymbol{h}}_{m}$ of all $\tilde{\boldsymbol{h}}_{I_i}^{'}$. 
Then, we calculate the distance between each $\tilde{\boldsymbol{h}}_{I_i}^{'}$ and $\tilde{\boldsymbol{h}}_{m}$ and update the value of $\tilde{\boldsymbol{h}}_{m}$ literally with the closest $\tilde{\boldsymbol{h}}_{I_i}^{'}$.
In this way, the mean value of the majority of $\tilde{\boldsymbol{h}}_{I_i}^{'}$ can be found, which is the estimated channel impulse response. By padding zeros and performing DFT, we can get the final estimated channel $\hat{\boldsymbol{H}}$.

\begin{algorithm}[t]
	\caption{1-Mean Filtering}
	\label{alg1}
	\begin{algorithmic}
		\STATE {\textbf{Inputs:} all estimated channel impulse response $\tilde{\boldsymbol{h}}_{I_i}^{'}$; Iteration Times $N_{iter}$}
		\STATE {\textbf{Outputs:} Estimated channel $\hat{\boldsymbol{H}}$};
		\STATE {\textbf{Initialization:} Compute the mean value of inputs:\\
		$\quad \tilde{\boldsymbol{h}}_{m} = \frac{\sum_{i=1}^{i=N_{1}}\tilde{\boldsymbol{h}}_{I_i}^{'}}{N_{1}}$;}
		\FOR{$i_{iter} \in [\![1, N_{iter}]\!]$}
		\STATE Compute the distance $d_{i}$ between each $\tilde{\boldsymbol{h}}_{I_i}^{'}$ and $\tilde{\boldsymbol{h}}_{m}$:
		\STATE $\quad d_{i} = \| \tilde{\boldsymbol{h}}_{I_i}^{'} - \tilde{\boldsymbol{h}}_{m} \|_2, i \in [\![1, N_{1}]\!]$;
		\STATE Select $\tilde{\boldsymbol{h}}_{I_i}^{'}$ with the top five minimum $d_{i}$;
		\STATE Update $\tilde{\boldsymbol{h}}_{m}$ by the mean value of the selected $\tilde{\boldsymbol{h}}_{I_i}^{'}$;
		\ENDFOR
		
		\STATE Padding zeros after $\tilde{\boldsymbol{h}}_{m}$ until its length equals the length of channel $N_c$:
		\STATE $\quad \hat{\boldsymbol{h}} = [\tilde{\boldsymbol{h}}_{m}, 0, 0, ..., 0] \in \mathbb{C}^{N_c \times 1};$
		\STATE Perform DFT to get the final estimated channel $\hat{\boldsymbol{H}}$:
		\STATE  $\quad \hat{\boldsymbol{H}} = \boldsymbol{F}\hat{\boldsymbol{h}}$;
	\end{algorithmic}
\end{algorithm}

\section{Performance Evaluation}\label{results}
\subsection{Simulation Setup}
In the simulation, we consider two multipath channels: Extended Vehicular A (EVA) model and Extended Typical Urban (ETU) model, which are defined in the 3GPP specification \cite{3gpp_36_104}. Table.~\ref{tab_channels} represents the delay and power of taps. The channel gains follow a Rayleigh fading and are subject to a circularly symmetric complex Gaussian (CSCG) distribution with zero mean and unit variance. The transmitted power is normalized to 1, and the noise power is varied to change the SNR condition. We set $N_1 = 10$.
And the time variation of the channel is simulated by the Markov chain model, the relation between CIRs of the last and current instants can be expressed as
\begin{align}
	\boldsymbol{h}_{t} = \lambda \boldsymbol{h}_{t-1} + \sqrt{1-\lambda^2}\boldsymbol{h}_{new}, \quad\lambda\in[0,1]
\end{align}
$\boldsymbol{h}_{new}$ is one channel realization totally independent to $h_{t-1}$. And $\lambda$ is the parameter describing the correlation of the channel in the dimension of time. When $\lambda=1$, the channel is static. When $\lambda=0$, the channels of two continuous instants are independent. The range of $\lambda$ in our simulation is set to $\{0.964, 0.984, 0.990, 0.999\}$.

For the payloads, each OFDM symbol contains 128 channels, and 14 OFDM symbols form one frame. The data symbols are modulated by QPSK.


We use the accuracy of hard decision $Acc = 1 - SER$ as the metric to represent the performance of different methods, where $SER$ implies the symbol error rate. The related parameter settings of RDDCE in the simulations are shown in  Table.~\ref{tab_setting}.

\begin{table}[b!]
	\caption{Channel parameters}
	\label{tab_channels}
	\centering
	\begin{tabular}{|c|ll|}
		\hline
		\multicolumn{1}{|l|}{\textbf{Types}} &
		\multicolumn{2}{c|}{\textbf{Parameters}}                                                         \\ \hline
		\multirow{2}{*}{EVA}               & \multicolumn{1}{l|}{Delay(ns)} & {[}0 30 150 310 370 710 1090 1730 2510{]}      \\ \cline{2-3} 
		& \multicolumn{1}{l|}{Power(dB)} & {[}0 -1.5 -1.4 -3.6 -0.6 -9.1 -7 -12 -16.9{]} \\ \hline
		\multirow{2}{*}{ETU}               & \multicolumn{1}{l|}{Delay(ns)} & {[}0 50 120 200 230 500 1600 2300 5000{]}      \\ \cline{2-3} 
		& \multicolumn{1}{l|}{Power(dB)} & {[}-1 -1 -1 0 0 0 -3 -5 -7{]}                  \\ \hline
	\end{tabular}
\end{table}
\begin{table}[b!]
	\caption{related parameters of RDDCE}
	\label{tab_setting}
	\begin{center}
		\begin{tabular}{|c|c|c|}
			\hline
			\textbf{Step} & \textbf{Parameter}& \textbf{Value} \\
			\hline
			\multirow{1}{*}{Channel Picking} & $N_{0}$ &  100 \\
			\cline{2-3}
			\hline 
			\multirow{2}{*}{DFT-based Channel Estimation} & $N_{1}$ &  15 \\
			\cline{2-3}
			& $N_{2}$ & 20 \\
			\hline 
			\multirow{1}{*}{1-Mean Filtering} & $N_{iter}$ &  10\\
			\cline{2-3}
			\hline
		\end{tabular}
	\end{center}
\end{table}

\subsection{Simulation Results} \label{sectionperformance}
\begin{figure}
	\centerline{\includegraphics[scale=0.6]{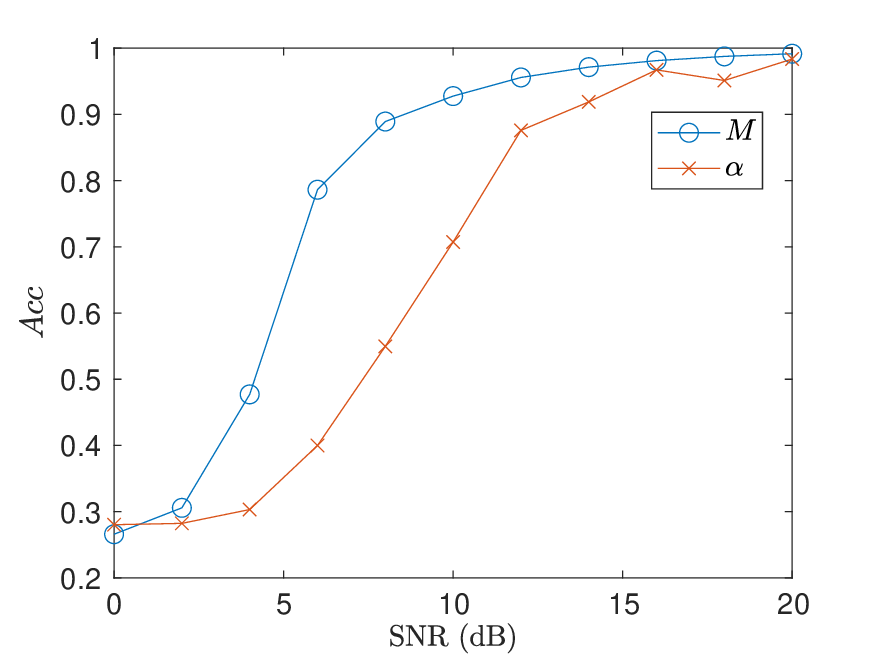}}
	\caption{$Acc$ of different metrics in terms of SNR in the channel selection step.}
	\label{metric}
\end{figure}

To show the effectiveness of the metric $M$ defined in Section~\ref{picking}, we have compared $Acc$ in terms of SNRs when applying $M$ and $\alpha$ in the channel selection stage. $\alpha$ is the metric used in \cite{fadingenvelope} to filter out the unreliable channels estimated by DDCE, which is defined as follows:
\begin{align}
	\alpha(t,k)= \|\hat{\boldsymbol{H}}_{t}[k]\|
\end{align}
As shown in Fig.~\ref{metric}, the metric $M$ outperforms $\alpha$, which demonstrates that $M$ is an effective metric that can filter out unreliable channels well. 

As a fair comparison, we also compare the $Acc$ of our method with the following four methods: ``Basic", ``Filtering", ``Interpolation" and ``Ideal". ``Basic" uses the preliminary estimation $\tilde{\boldsymbol{H}}_t$ as $\hat{\boldsymbol{H}}_t$ directly, which can be considered as the performance lower bound, ``Filtering" adapts the method proposed in \cite{gamma}, which updates the channel by a simple iterative filter. ``Interpolation" adapts the similar method like \cite{cooperativesubcarrier}, but all channels use the same modulation method. ``Ideal" uses the true channel $\boldsymbol{h}_{t,a}$ of every time instance $t$ to give the hard decision $\bar{\boldsymbol{X}}_t$, which can be considered as performance upper bound.

\begin{figure}
	\centerline{\includegraphics[scale=0.6]{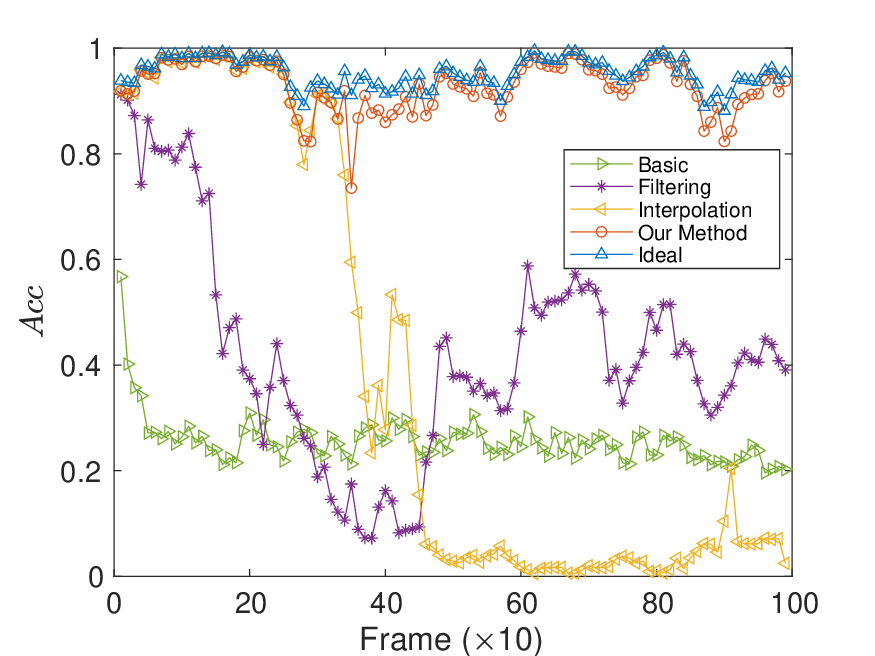}}
	\caption{$Acc$ of different methods on a long time scale.}
	\label{performance}
\end{figure}

We set SNR = 10 dB, and $\lambda=0.990$. The aforementioned methods are tested for 100 samples, each sample contains 1000 continuous frames. Fig.~\ref{performance} shows the variation of $Acc$ in one sample. We can observe that $Acc$ of ``Basic" reduces quickly and oscillates around 0.25, that is because we modulate symbols by QPSK in our simulation, which has four constellation points. After several frames, ``Basic" starts to decide $\bar{\boldsymbol{X}}_t$ randomly.
``Filtering" vibrates severely around 0.25.
The performance of ``Interpolation" relies on deciding $\bar{\boldsymbol{X}}_t$ at the chosen channels correctly. When the channel estimation error accumulates at these channels and exceeds the threshold of misjudgment occurring, its $Acc$ will reduce quickly.
Our RDDCE scheme chooses the reliable channels dynamically, which can avoid the performance degradation caused by error accumulation in the fixed channels, so it can keep high $Acc$ close to ``Ideal" on a long time scale.
The average $Acc$ of different methods for all samples are represented in Table.~\ref{tab_acc}. Our method outperforms other methods and has a close performance to ``Ideal".

\begin{table}[b!]
	\caption{$Acc$ for different channel models}
	\label{tab_acc}
	\centering
	\begin{tabular}{|c|c|c|c|c|c|}
		\hline
		\multicolumn{1}{|l|}{\multirow{2}{*}{\textbf{Types}}} &
		\multicolumn{5}{c|}{\textbf{Methods}}                                              \\ \cline{2-6}
		& Basic & Filtering & Interpolation & RDDCE & Ideal
		\\ \hline
		EVA & 0.264 & 0.292 & 0.519 & 0.917 & 0.949 
		\\ \hline
		ETU & 0.263 & 0.265 & 0.839 & 0.925 & 0.946
		\\ \hline
	\end{tabular}
\end{table}

\begin{figure}
	\centerline{\includegraphics[scale=0.6]{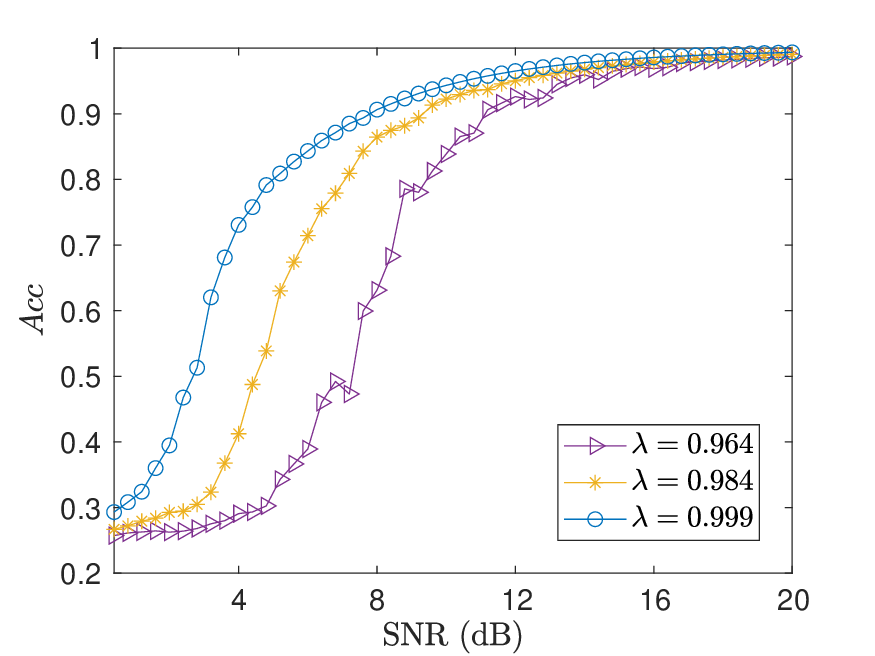}}
	\caption{$Acc$ of RDDCE in terms of SNR with different $\lambda$.}
	\label{snrlambdaacc}
\end{figure}

Noise and time variation of the channel are two important environmental factors that influence our method's performance, which are described by SNR and $\lambda$ in the simulation. We have compared the average $Acc$ of our method for 100 samples in terms of SNR with $\lambda \in \{0.964, 0.984, 0.999\}$, and each sample contains 1000 continuous frames. Fig.~\ref{snrlambdaacc} represents the result. In the range of SNR $\in [10,20]$, our method can effectively alleviate the performance degradation on a long time scale and so has high $Acc$. However, our method's performance degrades in the range of SNR $\in [0,10]$, especially for low $\lambda$. Finally, our method randomly decides $\bar{X}_t$ with equal probability like ``Basic".

Our RDDCE method exploits the channel correlation in time to update the channel for every instant.
Lower $\lambda$ reduces this correlation, causes more errors during demodulation, and finally makes it hard to select the subcarriers without misjudgment.
Lower SNR will also cause the misjudgment, besides it causes the estimation error itself.
Therefore, when $\lambda$ and SNR get small, our method cannot select enough subcarriers without misjudgment to compensate for the estimation error which leads to performance degradation.

\section{Conclusion}\label{conclusion}
In this paper, we propose the RDDCE method that combines conventional DDCE and DFT channel estimation to achieve reliable blind channel estimation on a long time scale.
By using $M$ and extended DFT channel estimation for subcarriers distributed unevenly, we select reliable subcarriers to estimate the channel and avoid the error accumulation caused by the misjudgment of hard decisions in conventional DDCE.
The simulation results show that compared with other DDCE methods, our method can achieve high-accuracy channel estimation for time-varying deep fading channels on a long time scale and has good performance to resist time variation and noise.

\bibliographystyle{IEEEtran}
\bibliography{ddce}

\vspace{12pt}

\end{document}